\documentclass[12pt]{llncs}

\usepackage[left=0.9in,top=0.9in,right=0.9in,bottom=0.9in]{geometry}
\usepackage{amsmath,amsfonts,amssymb,graphicx,epsfig} 
\usepackage{mathrsfs}

\usepackage[usenames,dvipsnames]{color}

\newcommand{\ent}{\ensuremath {\mathbb Z} }
\newcommand{\nat}{\ensuremath {\mathbb N} }

\newcommand{\pr} {{\bf Pr}}

\newcommand{\cD} {\ensuremath{\mathcal D}}

\newcommand{\remove}[1]{}

\newcommand{\cB}{\ensuremath {\mathcal B} }
\newcommand{\cL}{\ensuremath {\mathcal L} }

\newcommand{\eps}{\varepsilon}

\newcommand{\RG} {\ensuremath{\mathscr G}(n,r)}

\newcommand{\SR} {\ensuremath{\mathcal S}}

\def\wck#1 {\underline{#1}~\marginpar{\fbox{#1} {\tiny ?}}}
\def\silent#1\par{\par}
\def\text#1{\quad\mbox{#1}\quad}

\makeatletter
\renewcommand{\@seccntformat}[1]{\@nameuse{the#1}.\quad}
\makeatother

\DeclareMathOperator{\area}{Area}

\title{\Large{The domination number of on-line social networks and random geometric graphs}\thanks{Supported by grants from NSERC.}}
\author{Anthony Bonato \inst{1} \and Marc Lozier \inst{1} \and Dieter Mitsche \inst{2} \and \\  Xavier P\'{e}rez-Gim\'{e}nez \inst{1} \and Pawe\l\ Pra{\l}at \inst{1}}
\authorrunning{A. Bonato, M.\ Lozier, D. Mitsche, X.\ P\'{e}rez-Gim\'{e}nez, P.\ Pra{\l}at}
\institute{Ryerson University, Toronto, Canada, \\
\texttt{abonato@ryerson.ca}, \texttt{marc.lozier@ryerson.ca}, \\ \texttt{xperez@ryerson.ca}, \texttt{pralat@ryerson.ca}
\and
Universit\'{e} de Nice Sophia-Antipolis, Nice, France\\
\texttt{dmitsche@unice.fr}}

\date{}

\begin{document}

\maketitle

\begin{abstract} 
We consider the domination number for on-line social networks, both in a stochastic network model, and for real-world, networked data. Asymptotic sublinear bounds are rigorously derived for the domination number of graphs
 generated by the memoryless geometric protean random graph model. We establish sublinear bounds for the domination number of graphs in the Facebook 100 data set, and these bounds are well-correlated with those predicted
  by the stochastic model. In addition, we derive the asymptotic value of the domination number in classical random geometric graphs.
\end{abstract}

\section{Introduction}
On-line social networks (or OSNs) such as Facebook have emerged as a hot topic within the network science community. Several studies suggest OSNs satisfy many properties in common with other complex networks, such as:
power-law degree distributions~\cite{barabasi1999-scaling,Faloutsos-1999-power-law}, high local clustering~\cite{watts1998-dynamics}, constant~\cite{watts1998-dynamics} or even shrinking diameter with network
size~\cite{Leskovec-2007-densification}, densification~\cite{Leskovec-2007-densification}, and localized information flow bottlenecks~\cite{Estrada-2006-expansion,Leskovec-2009-community-structure}. Several models were
designed to simulate these properties~\cite{Kim-2012-mag,Kolda-2013-BTER}, and one model that rigorously asymptotically captures all these properties is the geometric protean model
(GEO-P)~\cite{Bonato-2012-geop,Bonato-2012-geop1,Bonato-2012-geop2} (see \cite{jp,lp,p,pw} for models where various ranking schemes were first used, and which inspired the GEO-P model). For a survey of OSN models see \cite{bon0}, and for more general complex networks \cite{bon1}. A fundamental difference with GEO-P versus other
models~\cite{barabasi1999-scaling,Kumar-2000-copying,Leskovec-2007-densification,Leskovec-2010-KronFit} is that it posits an underlying feature or metric space. This metric space mirrors a construction in the social
sciences called \emph{Blau space}~\cite{McPherson-1991-Blau}. In Blau space, agents in the social network correspond to points in a metric space, and the relative position of nodes follows the principle of
\emph{homophily}~\cite{McPherson-2001-homophily}: nodes with similar socio-demographics are closer together in the space. We give the precise definition of the GEO-P model (actually, one of its variants, the so-called
MGEO-P model) below. We focus on the MGEO-P model, since it simpler than GEO-P and generates graphs with similar properties.

The study of domination and dominating sets plays a prominent role in graph theory with a number of application to real-world networks. A \emph{dominating set} in a graph $G$ is a set of nodes $S$ in $G$ such that every
node not in $S$ is adjacent to at least one node in $S$. The \emph{domination number} of $G$, written $\gamma (G)$, is the minimum cardinality of a dominating set in $G$. Computing $\gamma (G)$ is a well-known
\textbf{NP}-complete problem, so typically heuristic algorithms are used to compute it for large-scale networks. Dominating sets appear in numerous applications such as: network controllability~\cite{cowan}, as a
centrality measure for efficient data routing~\cite{sto}, and detecting biologically significant proteins in protein-protein interaction network~\cite{m}. For more additional background on domination in graph theory,
see~\cite{Haynes}.

In social networks, we consider the hypothesis that minimum order dominating sets contain agents with strong influence over the rest of the network. Our goal in the present paper is to consider the problem of
finding bounds on dominating sets in stochastic models of OSNs, and also in real-world data derived from OSNs. We consider bounds on the domination number of a stochastic model (see next paragraph), and upper bounds for
that model are well-correlated with real-world OSN data. We note that the domination number has been studied previously in complex network models, including preferential attachment~\cite{cooper}, and recently
in~\cite{molnar}.

The OSN model we consider is called the \emph{memoryless geometric protean model} (\emph{MGEO-P)}, first introduced in~\cite{dim}. The MGEO-P model depends on five parameters which consists of: the number of nodes $n$,
the dimension of the metric space $m$, the attachment parameter $0 < \alpha < 1$, the density parameter $0 < \beta < 1-\alpha$, and the connection probability $0 < p \le 1.$

The nodes and edges of the network arise from the following process. Initially, the network is empty. At each of $n$ steps, a new node $v$ arrives and is assigned both a random position $q_v$ in $\mathbb{R}^m$ within
the unit-hypercube $[0,1)^m$ and a random rank $r_v$ from those unused ranks remaining in the set $1$ to $n$. The influence radius of any node $v$ is computed based on the formula:
\[ I(r_v) = \tfrac{1}{2} \bigl(r_v^{-\alpha} n^{-\beta}\bigr)^{1/m}. \] With probability $p$, the node $v$ is adjacent to an existing node $u$ satisfying $\mathcal{D}(v,u) \le I(r_u)$, where the distances are
computed with respect to the following metric: \[ \mathcal{D}(v,u) = \min\ensuremath{\left\{ \ensuremath{\left\| q_v-q_u -z \right\|_{\infty}} \colon z \in \ensuremath{\left\{ -1,0,1 \right\}}^m  \right\}},  \] and
where $\ensuremath{\left\| \cdot \right\|_{\infty}}$ is the infinity-norm. We note that this implies that the geometric space is symmetric in any point as the metric ``wraps'' around like on a torus. The volume of the
space influenced by the node is $r_v^{-\alpha} n^{-\beta}$. Then the next node arrives and repeats the process until all $n$ nodes have been placed. We refer to this model by MGEO-P$(n,m,\alpha,\beta,p)$.

We give rigorous bounds on the domination number of a typical graph generated by the MGEO-P. An event $A_n$ holds \emph{asymptotically almost surely} (\emph{a.a.s.}) if it holds with probability tending to
$1$ as $n$ tends to infinity. Our main result on MGEO-P is the following.

\begin{theorem}\label{main:geop}
If $m = o(\log n)$, then a.a.s.\ the domination number of a graph $G$ sampled from the MGEO-P$(n,m,\alpha,\beta,p)$ model satisfies
$$
\gamma(G) = \Omega(C^{-m/(1-\alpha)}n^{\alpha+\beta}) \text{ and } \gamma(G) = O(n^{\alpha+\beta}\log n),
$$
where $C$ is any constant greater than $6$. In particular, a.a.s.\ $\gamma(G) = n^{\alpha+\beta+o(1)}$.
\end{theorem}

We defer the proof of Theorem~\ref{main:geop} to Section~\ref{secone}. It is noteworthy that the domination number of the preferential attachment model is linear in the order of the graphs sampled; see~\cite{cooper};
this fact further demonstrates the differences between MGEO-P and other complex graph models.

Theorem~\ref{main:geop} suggests a sublinear bound on the domination number for OSNs, and we evidence for this in real-world data. In Section~\ref{sectwo}, we find bounds for graphs in the Facebook 100 data
set, and compare these results to those for the stochastic models. We chose to work with the so-called \emph{Facebook 100} (or FB100) data set, as it provides representative samples from the network of increasing orders.
Hence, we may consider trends for the domination number in the data. While the data presented is our first and initial study, the bounds we find for the domination number of FB100 are of sublinear order, and these bounds are well-correlated with those from MGEO-P.  Sublinear domination results for other complex networks were also reported in~\cite{molnar}; our approach is distinct as we consider social networks of increasing orders.

In addition to the results above, we find rigorous bounds on the domination number for classical random geometric graphs. Given a positive integer $n$, and a non-negative real $r$, we consider a \emph{random geometric
graph} $G=(V,E) \in\RG$ defined as follows. The node set $V$ of $G$ is obtained by choosing $n$ points independently and uniformly at random in the square $\SR = [0,1]^2$. (Note that, with probability $1$, no point in
$\SR$ is chosen more than once, and hence, we may assume that $|V|=n$.) For notational purposes, we identify each node $v \in V$ with its corresponding geometric position $v=(v_x,v_y)\in\SR$, where $v_x$ and $v_y$
denote the usual $x$- and $y$-coordinates in $\SR$, respectively. Finally, the edge set $E$ is constructed by connecting each pair of nodes $u$ and $v$ by an edge if and only if $d_E(u,v)\le r$, where $d_E$ denotes the
Euclidean distance in $\SR$.

Random geometric graphs were first introduced in a slightly different setting by Gilbert~\cite{Gilbert} to model the communications between radio stations. Since then several closely related variants on these graphs
have been widely used as a model for wireless communication, and have also been extensively  studied from a mathematical point of view. The basic reference on random geometric graphs is the monograph by
Penrose~\cite{Penrose}.

We note that our study is the first to explicitly provide provable bounds on the domination number of random geometric graphs. In particular, we derive the following result.
\begin{theorem}\label{thm:main}
Let $G\in\RG$ and let $\omega=\omega(n)$ be any function tending to infinity as $n \to \infty$. Then a.a.s.\ the following holds:
\begin{itemize}
\item [(a)] Denote by $N(x)$ the minimal number of balls of radius $x$ needed to cover $\SR$. If $r=\Theta(1)$, then
$$
\Omega(1) = N(r+\sqrt{\omega \log n /n}) \leq \gamma(G) \leq N(r-\omega/\sqrt{n}).
$$
\item [(b)] Define $C=2\pi\sqrt{3}/9 \approx 1.209.$ If $\omega \sqrt{\log n/n} \le r=o(1)$, then
$$
\gamma(G) =(C/\pi + o(1))r^{-2}.
$$
\item [(c)] If $1/\sqrt{n} \le r < \omega \sqrt{\log n/n}$, then
$$
\gamma(G) = \Theta(r^{-2}).
$$
\item [(d)] If $r < 1/\sqrt{n}$, then
$$
\gamma(G) = \Theta(n).
$$
\end{itemize}
\end{theorem}
The proof of Theorem~\ref{thm:main} is deferred to Section~\ref{secthree}. The final section summarizes our results and presents open problems.

\section{Proof of Theorem~\ref{main:geop}}\label{secone}

For each node $v\in[n]$, we consider the ball $B_v = \{ x\in[0,1)^m : \cD(x,v)\le I(r_v)\}$, which has volume $b_{v} = r_{v}^{-\alpha}n^{-\beta}$. The next lemma will be useful to estimate the sum of volumes of the
balls corresponding to a set of nodes.
\begin{lemma}\label{lem:sum}
Let $T$ be a set of $t$ nodes (fixed before ranks are chosen) with $\omega n^\alpha\log n \le t \le n$, for a function $\omega$ going to infinity with $n$ arbitrarily  slowly.
\begin{itemize}
\item[(a)] Then a.a.s.
\begin{equation}\label{eq:sum}
\sum_{i\in T} {r_i}^{-\alpha} = (1+o(1)) \frac{tn^{-\alpha}}{1-\alpha}.
\end{equation}
\item[(b)] Furthermore, given any integer $s$ such that $1\le s\le t$ and $s^{1-\alpha} \geq \omega (n/t)^{\alpha} \log n$, a.a.s.\ all subsets $S\subseteq T$ of $s$ nodes satisfy
\begin{equation}\label{eq:sum2}
\sum_{i\in S} {r_i}^{-\alpha} \le (1+o(1)) \frac{s^{1-\alpha} (t/n)^\alpha }{1-\alpha}.
\end{equation}
\end{itemize}
\end{lemma}
Observe that
the sum in~\eqref{eq:sum} is asymptotic to what one would expect. Indeed, if the ranks of the nodes in $T$ are distributed evenly, then one would obtain $\sum_{i\in T}^t {r_i}^{-\alpha} = \sum_{i=1}^t (in/t)^{-\alpha} =
tn^{-\alpha}/(1-\alpha) +O(1)$.
%
\begin{proof}
Let $s$ and $t$ be integers satisfying all the conditions of the statement in part (b).
Set $\hat\omega = \omega^{1/4}\to\infty$, so  we have $t\ge \hat\omega^4 (t/s)^{1-\alpha}n^\alpha\log n$. This also implies $\hat\omega=o((sn/t)^{1-\alpha})$.
Let $Y_j$ be the number of elements in $T$ with rank at most $j$. Observe that $Y_j$ has expectation $jt/n$, and follows  a hypergeometric distribution.
For $(sn/t)^{1-\alpha}/\hat\omega \le j\le n$, a Chernoff bound (see e.g.~\cite{JLR}) gives that
\[
\pr\left(\Big|Y_j-\frac{jt}{n}\Big|\ge (1/\hat\omega) \frac{jt}{n}\right) \le 2\exp\left(-\frac{jt}{3 \hat\omega^2 n}\right) \le 2 e^{-\hat\omega\log n/3} = o(1/n^2).
\]
We apply a union bound over all $j$, and conclude that a.a.s., for every $(sn/t)^{1-\alpha}/\hat\omega \le j\le n$,
\[
(1-1/\hat\omega)jt/n < Y_j < (1+1/\hat\omega)jt/n.
\]
In order to estimate the sums in the statement, we assume w.l.o.g.\ that $T=[t]$ and $r_1<r_2<\cdots<r_t$ (otherwise we permute the indices of the vertices in $T$). It follows that a.a.s., for every $(sn/t)^{1-\alpha}/\hat\omega \le j\le n$,
\[
r_{\lfloor (1-1/\hat\omega)jt/n\rfloor} \le j \le r_{\lceil(1+1/\hat\omega)jt/n\rceil}.
\]
Therefore, setting $\ell=2s^{1-\alpha}t^\alpha n^{-\alpha}/\hat\omega$, we have that a.a.s., for every $\ell \le i \le t$,
\[
\left\lfloor \frac{1}{1+1/\hat\omega}in/t \right\rfloor  \le r_ i \le \left\lceil \frac{1}{1-1/\hat\omega}in/t \right\rceil.
\]
For the lower bound on $r_i$ below, we need to use the fact that
$\left\lfloor \frac{1}{1+1/\hat\omega}in/t \right\rfloor \ge (sn/t)^{1-\alpha}/\hat\omega$, which is easily verified to be true since
$\hat\omega=o\left((sn/t)^{1-\alpha}\right)$. Finally, we infer that a.a.s., for any choice of $S$,
\[
\sum_{i\in S} {r_i}^{-\alpha} \le \sum_{i=1}^s {r_i}^{-\alpha}
= (1+o(1))\left(\frac{n}{t}\right)^{-\alpha} \sum_{i=\ell}^s i^{-\alpha} + O(\ell)
= (1+o(1)) \frac{1}{1-\alpha} s^{1-\alpha} t^\alpha n^{-\alpha}.
\]
This proves statement~(b). For statement~(a), take $s=t$ and note that for this choice of $s$, for any $\omega n^{\alpha} \log n \leq t \leq n$, the condition $s^{1-\alpha} \geq \omega (n/t)^{\alpha} \log n$ is satisfied. Observe that then $S=T=[t]$, so the first inequality in the above equation is an equality.\qed
\end{proof}
\subsubsection{Upper bound:}
Fix a constant $K>\frac{1-\alpha}{p}$, and let $D$ be the set containing the first $t=\lfloor K n^{\alpha+\beta}\log n\rfloor$ nodes added in the process.
We will show that a.a.s.\ $D$ is a dominating set. By Lemma~\ref{lem:sum}, we may condition on the event that~\eqref{eq:sum} holds for $t=|D|= \lfloor K
n^{\alpha+\beta}\log n \rfloor$. Note that this assumption on the ranks does not affect the distribution of the location of the nodes in $[0,1)^m$. Therefore, given a node $u>t$ (appearing in the process later than nodes in $D$), the probability that $u$ is not dominated by $D$ is
\begin{eqnarray*}
\prod_{i=1}^t(1-p {r_i}^{-\alpha} n^{-\beta}) &\le& \exp \left(- pn^{-\beta} \sum_{i=1}^t {r_i}^{-\alpha} \right) =
\exp\left(- (1+o(1)) \frac{p}{1-\alpha} tn^{-\alpha-\beta}\right) \\
&=& \exp\left(- (1+o(1)) \frac{Kp}{1-\alpha} \log n \right) = o(1/n).
\end{eqnarray*}
Taking a union bound over all nodes not in $D$, we can guarantee that a.a.s.\ all nodes are dominated.

As an alternative and relatively simple approach, one may prove the same upper bound on the domination number as follows. First, show that a.a.s.\ the minimum degree $\delta$ is at least $(1+o(1))pn^{1-\alpha-\beta}$.
Then we may use Theorem~1.2.2 in~\cite{as}, which states that for every graph $G$ with minimum degree $\delta$,
\begin{equation}\label{boundd}
\gamma (G) \le n\frac{1+\log(\delta+1)}{\delta+1}.
\end{equation}
\subsubsection{Lower bound:}
We consider for convenience a natural directed version of MGEO-P$(n,m,\alpha,\beta,p)$, by orienting each edge from its ``younger'' end node (that is, appearing later in the process) to its ``older'' end node. For a
set of nodes $D \subseteq[n]$, $N_{\mathrm{in}}(D)$ denotes the set of nodes $u \in [n] \setminus D$ such that there is a directed edge from $u$ to some node in $D$ or, equivalently, such that there is an edge from $u$
to some node in $D$ that is older than $u$. $N_{\mathrm{out}}(D)$ is defined analogously, replacing older by younger.

Define $t=\lceil n^{\alpha+\beta}\rceil$, and let $T=[t]$ be the set of the oldest $t$ nodes in the process. We want to show that a.a.s.\ there is no dominating set of order at most $\xi\mu^{-m}n^{\alpha+\beta}$, where
$\xi$ and $\mu$ are specified later. We give more power to our adversary by allowing her to pick (deterministically after the graph has been revealed) two sets of nodes $D_1$ and $D_2$ of order $\lfloor
\xi\mu^{-m}n^{\alpha+\beta} \rfloor$ each (not necessary disjoint). Her goal is also easier than the original one; she needs to achieve that, for every node $v\in T$, either $v$ is in-dominated by $D_1$ (that is, $v\in
D_1\cup N_{\mathrm{in}}(D_1)$) or $v$ is out-dominated by $D_2$ (that is, $v\in D_2\cup N_{\mathrm{out}}(D_2)$); nothing is required for young nodes in $[n] \setminus T$. We show that a.a.s.\ the adversary cannot
succeed, that is, regardless of her choice of $D_1,D_2$ we have always some node in $T$ not in $D_1\cup D_2\cup N_{\mathrm{in}}(D_1)\cup N_{\mathrm{out}}(D_2)$.
\paragraph{\underline{Out-domination:}}
Given a constant $0<\eps<1$, we define $T'$ to be the set of nodes in $T$ with rank greater than $(1-\eps)n$. Note that $|T'|$ has a hypergeometric distribution, so it follows easily from Chernoff's
bound (see~\cite{JLR}) that a.a.s.\ $|T'| \ge (\eps/2) n^{\alpha+\beta}$. For convenience, we choose $\eps=\eps(\alpha)$ to be the only real in $(0,1)$ satisfying
$\eps=2(1-\eps)^\alpha$. For every node $i\in T'$, the corresponding ball $B_i$ has length at most
\[
((1-\eps)^{-\alpha}n^{-\alpha-\beta})^{1/m} = ((2/\eps)n^{-\alpha-\beta})^{1/m}.
\]
We consider a tessellation of $[0,1)^2$ into large cells. At the centre of each large cell we consider a smaller cell. Small cells have side length $((2/\eps) n^{-\alpha-\beta})^{1/m}$ and large ones have side
length $2((2/\eps) n^{-\alpha-\beta})^{1/m}$. There are
\[
N = \Big\lfloor\frac{1}{2}((\eps/2) n^{\alpha+\beta})^{1/m}\Big\rfloor^m = \frac{\eps}{2} (2+o(1))^{-m}  n^{\alpha+\beta} \to\infty
\]
large cells fully contained in $[0,1)^m$ (we discard the rest), and thus $N$ small cells inside of those. By construction, if a node in $T'$ falls into a small cell, then its ball is contained into in the corresponding
large cell. Let $\mathcal X$ be the set of small cells that contain at least one node in $T'$, and let $T'' \subseteq T'$ be  a set of $X=|\mathcal X|$ nodes  such that each cell in $\mathcal X$ contains precisely one
node in $T''$ (if a given small cell contains at least two nodes in $T'$, then a node is selected arbitrarily to be placed in $T''$). Vertices in $T''$ are potentially dangerous for the adversary since, one node in
$D_2$ can ``out-dominate'' at most one single node in $T''$. However, she may in theory get lucky and in-dominate many of these nodes (in the next section we will show that this will not happen a.a.s.).

We want to show that a.a.s.\ $X\ge N/4$. The probability that there are at least $3N/4$ small cells containing no nodes in $T'$ is at most
\[
\binom{N}{\lceil 3N/4\rceil} \left(1-(3N/4) (2/\eps)n^{-\alpha-\beta}\right)^{(\eps/2)n^{\alpha+\beta}} \le
2^N \exp(-3N/4)=o(1).
\]
Therefore,
\begin{equation}\label{X}
X \ge N/4 = \frac{\eps}{8} \lambda^{-m} n^{\alpha+\beta}, \qquad \text{for some $\lambda=2+o(1)$.}
\end{equation}
\paragraph{\underline{In-domination:}}%
Let $\xi=\xi(\alpha)$ be a sufficiently small positive constant, and define $$\mu = \left( \lambda \left( 1+2(2/\eps)^{1/m}\right) \right)^{1/(1-\alpha)} > 3\lambda.$$ The adversary chooses a set $D_1\subseteq T$ of
$s= \lfloor\xi \mu^{-m}n^{\alpha+\beta} \rfloor$ nodes in her attempt to in-dominate $T'$. By Lemma~\ref{lem:sum}(b),  a.a.s.\ regardless of her choice,
\begin{equation}\label{eq:sum3}
\sum_{i\in D_1}{ r_i}^{-\alpha} \le  \tfrac{(1+o(1))}{1-\alpha} s^{-\alpha+1} t^\alpha  n^{-\alpha}= (1+o(1)) \tfrac{\xi^{1-\alpha}}{1-\alpha} \mu^{-(1-\alpha)m} n^{\beta}.
\end{equation}
We tessellate the space into cells of volume $(2/\eps) n^{-\alpha-\beta}$ (same size as the small cells in the out-domination part, but now we have the whole space partitioned into cells of that size). Recall that,
for each node $i\in D_1$, the ball $B_i$ has length ${b_i}^{1/m}\ge n^{-(\alpha+\beta)/m}$. Therefore, the volume of the set of cells intersected by $B_i$  is at most
\[
\left({b_i}^{1/m} + 2\big( (2/\eps) n^{-\alpha-\beta}\big)^{1/m} \right)^m \le \left(1 + 2 (2/\eps)^{1/m} \right)^m b_i
= \left( \mu^{1-\alpha}/\lambda \right)^m b_i.
\]
Combining this and~\eqref{eq:sum3}, a.a.s.\ and regardless of the adversary's choice, the total volume of the cells intersected by the balls of the nodes in $D_1$ is at most
\begin{equation} \label{aaa}
\sum_{i\in D_1} \left( \mu^{1-\alpha}/\lambda \right)^m b_i = \left( \mu^{1-\alpha}/\lambda \right)^m n^{-\beta} \sum_{i\in D_1} {r_i}^{-\alpha}
\le (1+o(1)) \tfrac{\xi^{1-\alpha}}{1-\alpha} \lambda^{-m}.
\end{equation}
Let $\mathcal Y$ be the set of cells  intersected by the balls of the nodes in $D_1$, and put $Y=|\mathcal Y|$. By~\eqref{aaa}, a.a.s.\
\[
Y \le \frac{\eps \xi^{1-\alpha}}{2(1-\alpha)} \lambda^{-m} n^{\alpha+\beta}.
\]
Thus, in view of~\eqref{X} we just need to make $\xi$ small enough so that $|\mathcal X\setminus\mathcal Y|$ is larger than $|D_2|=\lfloor \xi \mu^{-m} n^{\alpha+\beta}\rfloor$. That is because dangerous cells in
$\mathcal X\setminus\mathcal Y$ contain nodes in $T'$ that are not in-dominated by $D_1$, and each one of these cells requires one different node in $D_2$ to out-dominate its nodes. Recall that our choice of
$\eps\in(0,1)$ depends only on $\alpha$. Then picking $\xi$ sufficiently small so that $\frac{\eps \xi^{1-\alpha}}{2(1-\alpha)} + \xi < \frac{\eps}{8}$, we get
\[
|\mathcal X\setminus\mathcal Y| \ge X-Y \ge
\left(\frac{\eps}{8} - \frac{\eps \xi^{1-\alpha}}{2(1-\alpha)} \right) \lambda^{-m} n^{\alpha+\beta}
> \xi \mu^{-m} n^{\alpha+\beta} \ge \lfloor \xi \mu^{-m} n^{\alpha+\beta} \rfloor = |D_2|,
\]
where we also used that $\mu>3\lambda>\lambda$. Finally, distinguishing the cases $m=O(1)$ and $m\to\infty$, we observe that
\[
\mu^{-m} = \left(\lambda(1+2(2/\eps)^{1/m})\right)^{-m/(1-\alpha)}
=
\begin{cases}
\Theta(1) & \text{for $m=O(1)$,}
\\
(6+o(1))^{-m/(1-\alpha)}  & \text{for $m\to\infty$,}
\end{cases}
\]
so the claimed lower bound follows.
\section{Domination in Facebook 100 graphs}\label{sectwo}

Facebook distributed 100 samples of social networks from universities within the United States measured as of September 2005~\cite{Traud-2011-facebook}, which range in size from 700 nodes to 42,000 nodes. We call these
networks the \emph{Facebook 100} (or simply FB100) graphs. As the domination number is sensitive to nodes of low degree, we used the $k$-core of the network, where $1\le k \le 5$; see~\cite{Seidman1983-cores}. For $k
\in \nat,$ the \emph{$k$-core} of a graph is the largest induced subgraph of minimum degree at least $k$. The $k$-core can be found by a simple node deletion algorithm that repeatedly deletes nodes with degree less than
$k$. This algorithm always terminates with the $k$-core of the graph, which is possibly empty.

Several algorithms were used to bound the domination number of the FB100 graphs, but one providing the smallest dominating sets is an adaptation of the \emph{DS-DC} algorithm \cite{m}. In the algorithm, initially all nodes $V$
are in the dominating set $S$. It then selects a node $u$ of minimum degree in $S$, and deletes it only if the set $S \setminus \{u \}$ remains dominating. The algorithm then repeats these steps for all nodes in $S$ in
order of their increasing degrees. We considered other algorithms, such as greedy algorithms where high degree nodes are added to an empty dominating set sequentially, or by choosing a random dominating set, but DS-DC
outperformed these algorithms. We omit a detailed discussion of the performance of other algorithms owing to space.

\begin{figure} [h!]
\begin{center}
\epsfig{figure=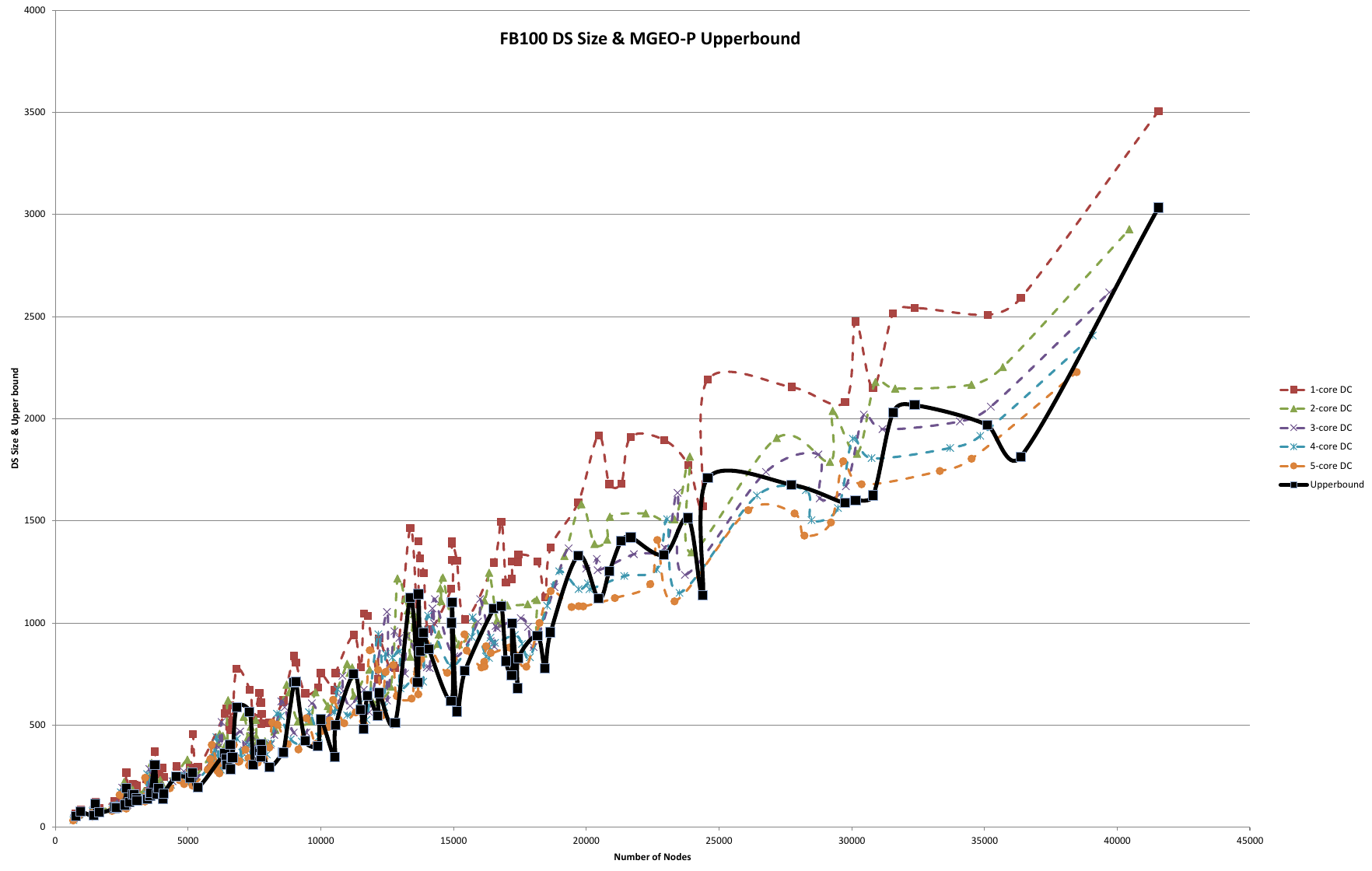,scale=0.7}
\caption{Upper bounds on the domination number of the FB100 networks vs MGEO-P.}\label{fig1}
\end{center}
\end{figure}

Figure~\ref{fig1} presents the DS-DC predicted upper bounds on $\gamma(G)$, where $G$ is a graph in the FB100 data set.
We plotted the upper bound predicted by the MGEO-P model in Theorem~\ref{main:geop}, and we note the close similarity between that bound and ones for FB100. Note that we ignore constants in the big Oh term in the upper bound
from the model, and simply plot the bound generated by $n^{\alpha + \beta} \log n$. The values for $\alpha$, $\beta$, and the dimension parameter $m$ for each of the FB100 graphs are taken from tables provided
in~\cite{dim}. (For example, in order to determine the power-law exponent, the Clauset-Shalizi-Newman power law exponent estimator was used; see~\cite{dim} for more details.) The MGEO-P bound seems well-correlated with the bounds provided in the $k$-core, especially where $k=3,4,5.$ See Table~\ref{table:1}, which fits the domination number of the FB100 graphs to the curve $y=n^x\log n$.
\begin{table}[h!]
\begin{center}
\begin{tabular}{|c|c|c|}
\hline
$k$ & $x$ & $R^{2}$ \\ \hline\hline
$1$ & \multicolumn{1}{||l|}{$0.509$} & $0.8472$ \\ \hline
$2$ & \multicolumn{1}{||l|}{$0.492$} & $0.8292$ \\ \hline
$3$ & \multicolumn{1}{||l|}{$0.4818$} & $0.8179$ \\ \hline
$4$ & \multicolumn{1}{||l|}{$0.4741$} & $0.8093$ \\ \hline
$5$ & \multicolumn{1}{||l|}{$0.4677$} & $0.803$ \\ \hline
\end{tabular}
\smallskip 	
\caption{Fitting the domination number of the $k$-cores of FB100 to $y=n^x\log n$.}\label{table:1}
\end{center}
\end{table}

\begin{figure} [h!]
\begin{center}
\epsfig{figure=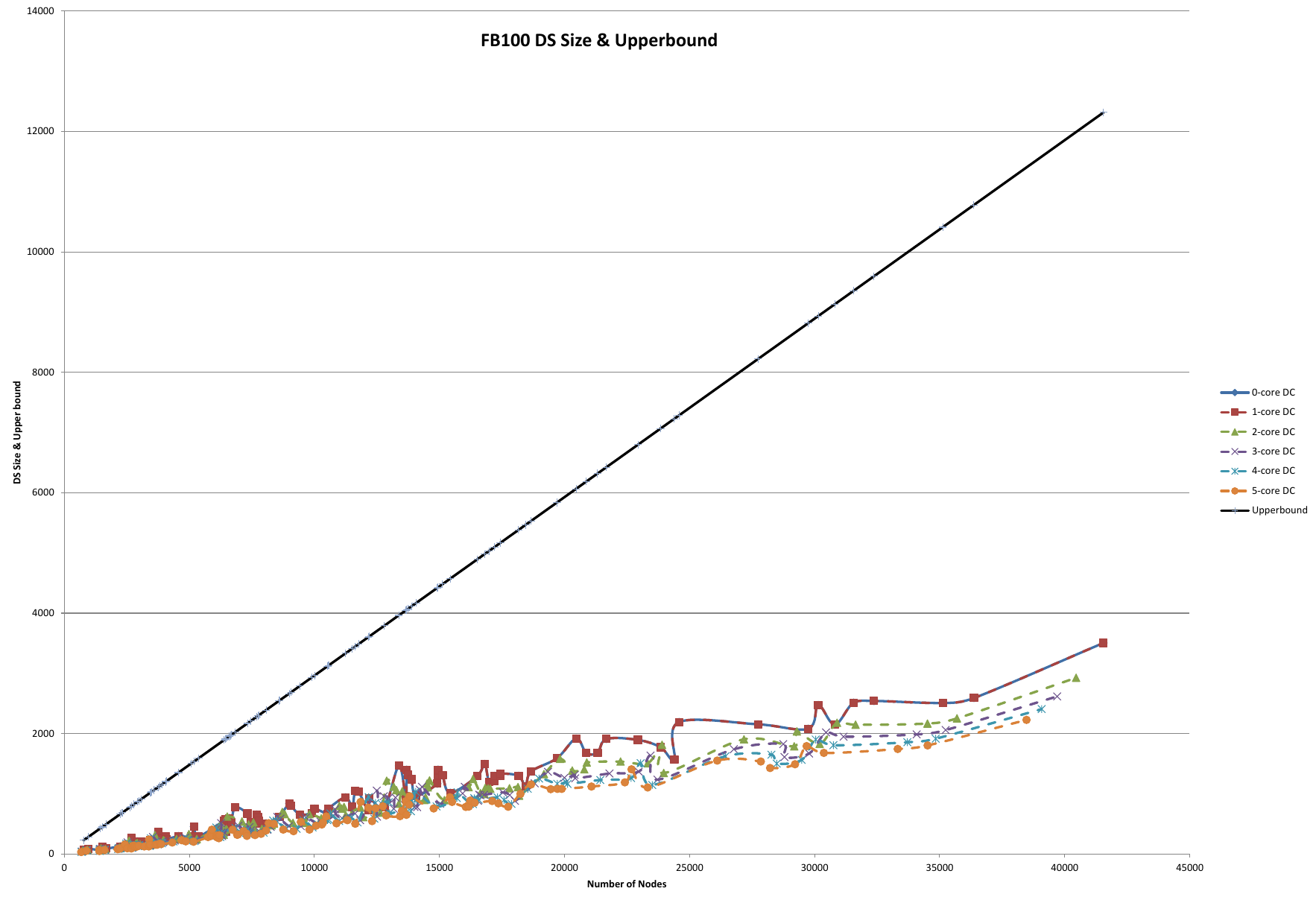,scale=0.6}
\caption{Upper bounds on the domination number of the FB100 networks vs the bound in (\ref{boundd})}\label{fig2}
\end{center}
\end{figure}

To contrast the bounds provided in Figure~\ref{fig1} with the bound in (\ref{boundd}), we plot them in Figure~\ref{fig2}. We plotted the theoretical bound using $\delta = 5$ (that is, the minimum degree of the
$5$-core).
The figure shows a significant over-estimate of the domination number of the bound in (\ref{boundd}), further corroborating the claim that the domination numbers of the FB100 graphs are sublinear with respect to the
order of the graph.

\section{Proof of Theorem~\ref{thm:main}}\label{secthree}

 We will relate the domination number to the problem of covering the plane with circles. Given $x\in\mathbb{R}^2$ and $\rho>0$, we denote by $\cB(x,\rho)$ the ball with centre $x$ and of radius $\rho$. The following theorem is well known~\cite{Kershner}.

\begin{theorem}[\cite{Kershner}]\label{thm:kershner}
Given a bounded subset of the plane $M$, for $\varepsilon  > 0$ let $N(\varepsilon)$ be the minimum number of balls of radius $\varepsilon$  that can cover $M$. Then we have that
$$
\lim_{\varepsilon \to 0} \pi \varepsilon^2 N(\varepsilon) = C \area(\overline{M}),
$$
where $\overline{M}$ denotes the closure of $M$.
\end{theorem}

Observe that $(C-1)$ can therefore, be seen as measuring the proportion of unavoidable overlapping. Moreover,~\cite{Kershner} shows that an optimal covering of the square $\SR$ using balls of radius $\varepsilon$
corresponds to arranging the balls in such a way that their centers are the centers of the cells of a hexagonal tiling of length $\varepsilon$. More precisely, consider the lattice
\begin{equation}\label{eq:lattice}
\cL_\eps= \{i\eps(\sqrt 3,0) + j\eps(\sqrt 3/2,3/2) : i,j\in\ent\}.
\end{equation}
Then the set of balls of radius $\eps$ and centre in $\cL_\eps$ that intersect $\SR$ form a covering of $\SR$ that gives the limit in Theorem~\ref{thm:kershner}.

Note that for all $G$ with maximum degree $\Delta$, we trivially have $\gamma(G) \geq n/(1+\Delta(G))$ (for further relations between $\gamma(G)$ and other graph parameters, see, for example,~\cite{Haynes}). Given any
constant $c>0$, for $G \in \RG$ with $r \geq c \sqrt{\log n/n}$, it is easy to show, by Chernoff bounds together with union bounds, that a.a.s.\ $\Delta(G) = O(r^2 n)$. Therefore, a.a.s.\ we have
$\gamma(G)=\Omega(r^{-2})$. On the other hand, we can trivially construct a dominating set of $\RG$ by tessellating $\SR$ into square cells of side length $r/\sqrt2$ and picking one node from each cell (if the cell is
not empty). This holds deterministically for any geometric graph (not necessarily random), with no restriction on $r$, and gives $\gamma(G)=O(r^{-2})$. It follows that, for $G \in \RG$ with $r \geq c \sqrt{\log n/n}$,
a.a.s.\ $\gamma(G) = \Theta(r^{-2})$.

\smallskip

We first prove the lower bound in part (b). Fix an arbitrarily small constant $\delta > 0$. Tessellate $\SR$ into cells of side length $\alpha=\sqrt{\omega \log n/n}=o(r)$. By Chernoff bounds together with a union bound
over all cells, we get that a.a.s.\ each cell contains at least one node. We may condition on this event, and proceed deterministically. For a contradiction, suppose that there exists a dominating set of size
$s=\lfloor(C/\pi-\delta)r^{-2} \rfloor.$ Consider then $s$ balls whose centers are at the nodes of the dominating set.  When using radius $r$ for all balls, each cell is at least touched by some ball, since each cell is
non-empty and each node is covered. Hence, by using radius $r'=r+\alpha \sqrt{2}$, each square is totally covered by some ball. Therefore, $\SR$ can be  covered by $\lfloor(C/\pi-\delta)r^{-2}\rfloor$ balls of radius
$r'$. On the other hand,
$$
\pi r'^2 \lfloor(C/\pi-\delta) r^{-2} \rfloor \le (r^2+2\sqrt{2}r\alpha + 2\alpha^2) (C-\pi\delta) r^{-2} = (1+o(1)) (C-\pi\delta),
$$
since $\alpha = o(r)$. This contradicts Theorem~\ref{thm:kershner} and so $\gamma(G) > s$. Since the argument holds for any $\delta > 0$, we get the desired lower bound.

For the upper bound, we will show that we can find a covering of $\SR$ with $(C/\pi + o(1))r^{-2}$ balls of radius $r$ that are centered at some nodes of $G$. Again, fix some arbitrarily small constant $\delta > 0$. Let
$r'=(1-\delta)r$, and consider the lattice $\cL_{r'}$, as defined in~\eqref{eq:lattice}. Let $\cL'_{r'}$ be the set of all points $x\in\cL_{r'}$ such that the ball with centre $x$ and radius $r'$ intersects $\SR$.
Recall that $\cL'_{r'}$ gives the optimal covering of $\SR$ with balls of radius $r'$, and therefore, attains the bound given by Theorem~\ref{thm:kershner}
$$
s= |\cL'_{r'}| = \left( \frac{C}{\pi r'^{2}} \right)(1+o(1))= \left( \frac {C}{\pi} \right)r^{-2} (1-\delta)^{-2}(1+o(1)).
$$
It might happen that some point $x\in\cL'_{r'}$ does not belong to $\SR$. In this case, we replace $x$ by the closest point $\hat x$ on the boundary of $\SR$ (this can be uniquely done, since $\SR$ is closed and convex).
Note that $\cB(x,r')\cap\SR\subseteq\cB(\hat x,r')\cap\SR$. We denote $\hat\cL_{r'}$ the modified set of points that we obtained. By construction, $\hat\cL_{r'}\subseteq\SR$, and we can cover $\SR$ using balls with centre
 in $\hat\cL_{r'}$ and radius $r'$ or larger. Moreover, $|\hat\cL_{r'}| = s$. Clearly, if we can guarantee that for each $x\in\hat\cL_{r'}$ there exists a node of $G$ inside $\mathcal{B}(x,\delta r) \cap \SR$, then
 $G$ is dominated by these nodes, and hence, $s$ yields an upper bound for $\gamma(G)$.

Observe that for any point $x\in\hat\cL_{r'}$ (and therefore, in $\SR$), the area of $\mathcal{B}(x,\delta r) \cap \SR$ is at least $(\delta r)^2 \pi / 4$, since at least a quarter of a ball must be inside $\SR$. The
probability that there is no node of $G$ in $\mathcal{B}(x,\delta r) \cap \SR$ is at most
$$
\left( 1-\frac{(\delta r)^2 \pi }{4} \right)^n \leq \exp \left( - \frac {n(\delta r)^2 \pi}{4} \right)  \le \exp \left( - \frac {\omega^2 \delta^2 \pi \log n }{4} \right) = o(n^{-2}) .
$$
Since there are $s$ events that we need to investigate and clearly $s \le n$, by a union bound, a.a.s., for every $x\in\hat\cL_{r'}$, the region $\mathcal{B}(x,\delta r) \cap \SR$ contains at least one node of $G$.  It
follows that a.a.s.\ $\gamma(G) \le s$ and since the argument holds for any $\delta > 0$, we derive the desired upper bound.

For the proof of part (a), note that $N(x)$ is non-decreasing function of $x$, and $N(x) =1$ for $x \geq 1/\sqrt{2}$. Fix $r=\Theta(1)$. Tessellate $\SR$ into cells of side length $\alpha=\sqrt{(\omega/2) \log n /n}$.
For the lower bound, suppose for contradiction that $\gamma(G) \leq N(r+\alpha \sqrt{2})-1$.  By Chernoff bounds together with a union bound over all cells, a.a.s.\ there is at least one node in each such cell. Now
place $N(r+\alpha \sqrt{2})-1$ many balls with centers at the nodes of the dominating set. Since by using radius $r$ each cell is at least touched by some ball, by using radius $r+\alpha \sqrt{2}$ each cell is totally
covered by a ball, and therefore $\SR$ is covered by $N(r+\alpha \sqrt{2})-1$  balls of radius $r+\alpha \sqrt{2}$, contradicting the definition of $N(x)$. Therefore,  a.a.s.\ $\gamma(G) \geq N(r+\alpha \sqrt{2})$.

For the upper bound, consider an optimal arrangement of $N(r-\beta)$ balls of radius $r-\beta$, where $\beta=\omega/\sqrt{n}$. As before, if the centre $p$ of a ball is outside $\SR$, but
$\cB(p,r-\beta)\cap\SR\ne\emptyset$, we may shift the centre of the ball towards its closest point $p'$ on the boundary of $\SR$. Since $\cB(p,r-\beta)\cap\SR\subseteq\cB(p',r-\beta)\cap\SR$, we still preserve the
covering property, and therefore, we can obtain an optimal covering of $\SR$ with balls of radius $r-\beta$ and centered at points inside of $\SR$. As in part (a), it suffices to show the existence of a node $v \in V$
inside $\mathcal{B}(c,\beta) \cap \SR$ for any centre $c$ in this optimal arrangement of balls. Since $N(r-\beta) = O(1)$, the probability that there exists a centre $c$ such that $(\mathcal{B}(c,\beta) \cap \SR) \cap V
= \emptyset$ is at most
$$
O(1) (1-\beta^2 \pi / 4)^n = O(\exp(-n \beta^2\pi /4))=o(1),
$$
and hence, a.a.s.\ for all centers $c$, we have that $\mathcal{B}(c,\beta) \cap \SR$ contains at least one node of $G$. These nodes form a dominating set, and so a.a.s.\ $\gamma(G) \leq N(r-\beta)$.

Finally, note that the lower bound in part (b) can be easily adopted to show that a.a.s.\ $\gamma(G) = \Omega(r^{-2})$ as a.a.s.\ a positive fraction of cells contain at least one node for the range of $r$ considered in
part (c). As already mentioned, the upper bound of $O(r^{-2})$ holds for any (deterministic) geometric graph and any $r$. Hence, part (c) follows. For part (d), the upper bound is trivial. The lower bound comes from the
fact that a.a.s. there will be $\Theta(n)$ isolated nodes, and a dominating set has to contain all of them. The proof of the theorem is finished. \qed

\section{Conclusions and open problems}

We considered the domination number of a stochastic model for OSNs, the MGEO-P model. Theorem~\ref{thm:main} shows a sublinear bound on the domination number of OSNs, which is well correlated with estimates for the
domination number taken for the Facebook 100 data set. In addition, we provided bounds for the domination number of random geometric graphs.

In future work, we would like to broaden our analysis of the domination number to other data sets, and to test larger samples of OSNs. We will contrast the estimates provided by other heuristic algorithms for computing
minimum order dominating sets, and provide a fitting of the data to bounds provided by the model.

So-called ``elites'', those who exert strong influence on the ambient network, are studied extensively in the sociology literature (see~\cite{elites} for an overview of the literature on this topic). One approach to
detecting elites is via their relatively high degree; hence, the use of $k$-cores in~\cite{elites}. A different approach to detecting elites is to search for them within a minimum order dominating set, as these sets
reach the entire network. Further, if minimum order dominating sets have much smaller order than the network (as we postulate), then that reduces the computational costs of finding elites. We plan on considering this approach to finding elites via dominating sets in future work.


\begin{thebibliography}{99}

\bibitem{as} N.\ Alon, J.\ Spencer, \emph{The Probabilistic Method}, Wiley,
New York, 2000.

\bibitem{barabasi1999-scaling}
A.L.\ Barab{\'a}si, R.\ Albert, Emergence of scaling in random networks, \emph{Science} \textbf{286} (1999) 509--512.

\bibitem{bon1} A.~Bonato, \emph{A Course on the Web Graph}, American Mathematical Society Graduate Studies Series in Mathematics, {Providence}, Rhode Island, 2008.

\bibitem{dim} A.\ Bonato, D.F.\ Gleich, M.\ Kim, D.\ Mitsche, P.\ Pra\l{}at, A.\ Tian, S.J.\ Young, Dimensionality matching of social networks using motifs and eigenvalues, \emph{PLOS ONE} \textbf{9} (2014) e106052.

\bibitem{Bonato-2012-geop}
A.\ Bonato, J.\ Janssen, P.\ Pra{\l}at, Geometric protean graphs, \emph{Internet Mathematics} \textbf{8} (2012) 2--28.

\bibitem{Bonato-2012-geop1} A.\ Bonato, J.\ Janssen, P.\ Pra{\l}at, The geometric protean model for on-line social networks, In: \emph{Proceedings of Workshop on Algorithms and Models for the Web Graph (WAW'10)},
    2010.

\bibitem{Bonato-2012-geop2} A.\ Bonato, J.\ Janssen, P.\ Pra{\l}at, A geometric model for on-line social networks, In: \emph{Proceedings of 3rd Workshop on Online Social Networks (WOSN'10)}, 2010.

\bibitem{bon0} A.\ Bonato, A.\ Tian, Complex networks and social networks, invited book chapter in: \emph{Social Networks}, editor E. Kranakis, Springer, Mathematics in Industry series, pp.\ 269-285, 2013.

\bibitem{cooper} C.\ Cooper, R.\ Klasing, M.\ Zito, Lower bounds and algorithms for
dominating sets in web graphs, \emph{Internet Mathematics} \textbf{2} (2005) 275–-300.

\bibitem{elites} B.\ Corominas-Murtra, B.\ Fuchs, S.\ Thurner, Detection of the elite structure in a virtual multiplex social system by means of a generalized $k$-core, Preprint 2014.

\bibitem{cowan} N.J.\ Cowan, E.J.\ Chastain, D.A.\ Vilhena, J.S.\ Freudenberg, C.T.\ Bergstrom, Nodal dynamics, not degree distributions, determine the structural controllability of complex networks, \emph{PLOS ONE}
\textbf{7} (2012) e38398.

\bibitem{Estrada-2006-expansion} E.\ Estrada, Spectral scaling and good expansion properties in complex networks, \emph{Europhysics Letters} \textbf{73} (2006) 649.

\bibitem{Faloutsos-1999-power-law}
M.\ Faloutsos, P.\ Faloutsos, C.\ Faloutsos, On power-law relationships of the   internet topology, \emph{SIGCOMM Comput Commun Rev} (1999) \textbf{29} 251--262.

\bibitem{Gilbert} E.N.~Gilbert, Random plane networks, {\em  J. Soc. Industrial Applied Mathematics} \textbf{9} (1961) 533--543.

\bibitem{Haynes} T.W.~Haynes, S.T.~Hedetniemi, P.J.~Slater, {\em Fundamentals of Domination in Graphs}, CRC Press, 1998.

\bibitem{jp} J.\ Janssen, P.\ Pra{\l}at, Protean graphs with a variety of ranking schemes, \emph{Theoretical Computer Science} \textbf{410} (2009) 5491--5504.

\bibitem{JLR} S.\ Janson, T.\ \L uczak and A.\ Rucinski, {\em Random Graphs}, Wiley-Intersci.\ Ser.\ Discrete Math.\ Optim., 2000.

\bibitem{Kershner}
R.~Kershner, The number of circles covering a set, {\em American Journal of Mathematics} \textbf{61} (1939) 665--671.

\bibitem{Kim-2012-mag}
M.\ Kim, J.\ Leskovec, Multiplicative attribute graph model of real-world networks, \emph{Internet Mathematics} \textbf{8} (2012) 113--160.

\bibitem{Kolda-2013-BTER}
T.G.\ Kolda, A.\ Pinar, T.\ Plantenga, C.\ Seshadhri, A scalable generative graph model with community structure, Preprint 2014.

\bibitem{Kumar-2000-copying}
R.\ Kumar, P.\ Raghavan, S.\ Rajagopalan, S.\ Sivakumar, A.\ Tomkins,
  Stochastic models for the web graph, In: \emph{Proceedings of the 41st Annual Symposium on Foundations of
  Computer Science}, 2000.

\bibitem{Leskovec-2010-KronFit}
J.\ Leskovec, D.\ Chakrabarti, J.\ Kleinberg, C.\ Faloutsos, Z.\ Ghahramani, Kronecker graphs: An approach to modeling networks, \emph{Journal of Machine Learning Research} \textbf{11} (2010) 985--1042.

\bibitem{Leskovec-2007-densification}
J.\ Leskovec, J.\ Kleinberg, C.\ Faloutsos, Graph evolution: Densification and
  shrinking diameters, \emph{ACM Trans Knowl Discov Data} \textbf{1} (2007) 1--41.

\bibitem{Leskovec-2009-community-structure}
J.\ Leskovec, K.J.\ Lang, A.\ Dasgupta, M.W.\ Mahoney, Community structure in large
  networks: Natural cluster sizes and the absence of large well-defined
  clusters, \emph{Internet Mathematics} \textbf{6} (2009) 29--123.

\bibitem{lp} T.\ Luczak, P.\ Pra{\l}at, Protean graphs, \emph{Internet Mathematics} \textbf{3} (2006) 21--40.

\bibitem{McPherson-1991-Blau}
J.M.\ McPherson, J.R.\ Ranger-Moore, Evolution on a dancing landscape:
  Organizations and networks in dynamic blau space, \emph{Social Forces} \textbf{70} (1991) 19--42.

\bibitem{McPherson-2001-homophily}
M.\ McPherson, L.\ Smith-Lovin, J.M.\ Birds of a feather: Homophily in
  social networks, \emph{Annual Review of Sociology} \textbf{27} (2001) 415--444.

\bibitem{m}
T.\ Milenkovi\'{c}, V Memi\v{s}evi\'{c}, A.\ Bonato, N.\ Pr\v{z}ulj, Dominating biological networks,
\emph{PLOS ONE} \textbf{6} (2013) (8), e23016.

\bibitem{molnar} F.\ Moln\'{a}r Jr., N.\ Derzsy, \'{E}. Czabarka, L.\ Sz\'{e}kely, B.K.\ Szymanski, G.\ Korniss,
Dominating scale-free networks using generalized probabilistic methods, Preprint 2014.

\bibitem{p} P.\ Pra{\l}at, A note on the diameter of protean graphs, \emph{Discrete Mathematics} \textbf{308} (2008) 3399--3406.

\bibitem{pw} P.\ Pra{\l}at, N.\ Wormald, Growing protean graphs, \emph{Internet Mathematics} \textbf{4} (2009) 1--16.

\bibitem{Penrose}
M.~Penrose, {\em Random Geometric Graphs}, Oxford Studies in Probability. Oxford U.P., 2003.

\bibitem{Seidman1983-cores}
S.B.\ Seidman, Network structure and minimum degree, \emph{Social Networks} \textbf{5} (1983) 269--287.

\bibitem{sto}
I.\ Stojmenovic, M.\ Seddigh, J.\ Zunic, Dominating sets and neighbor
elimination-based broadcasting algorithms in wireless networks, \emph{IEEE Transactions on Parallel and Distributed Systems} \textbf{13} (2002) 14-–25.

\bibitem{Traud-2011-facebook}
A.L.\ Traud, P.J.\ Mucha, M.A.\ Porter, Social structure of Facebook networks, Preprint 2014.

\bibitem{watts1998-dynamics}
D.J.\ Watts, S.H.\ Strogatz, Collective dynamics of ``small-world'' networks, \emph{Nature} \textbf{393} (1998) 440-442.

\bibitem{west} D.B.\ West, \emph{Introduction to Graph Theory, 2nd edition},
Prentice Hall, 2001.

\bibitem{Zhang-2006-Apollonian}
Z.\ Zhang, F.\ Comellas, G.\ Fertin, L.\ Rong, High-dimensional apollonian networks, Journal of Physics A: Mathematical and General (2006) \textbf{39} 1811.

\end{thebibliography}
\end{document}